\documentclass[aps,prd,twocolumn,nofootinbib,superscriptaddress]{revtex4-2}

\usepackage{amsmath,amssymb,graphicx,hyperref}
\usepackage{bm}

\begin{document}

\title{Electroweak Doublet Dark Matter for a Galactic Halo Gamma-Ray Excess}

\author{Yasunori Nomura}
\affiliation{Leinweber Institute for Theoretical Physics, Department of Physics, University of California, Berkeley, CA 94720, USA}
\affiliation{Theoretical Physics Group, Lawrence Berkeley National Laboratory, Berkeley, CA 94720, USA}
\affiliation{RIKEN Center for Interdisciplinary Theoretical and Mathematical Sciences (iTHEMS), RIKEN, Wako 351-0198, Japan}
\affiliation{Kavli Institute for the Physics and Mathematics of the Universe (WPI), UTIAS, The University of Tokyo, Kashiwa, Chiba 277-8583, Japan}

\author{Tomonori Totani}
\affiliation{Department of Astronomy, School of Science, The University of Tokyo, Bunkyo-ku, Tokyo 113-0033, Japan}
\affiliation{Research Center for the Early Universe, School of Science, The University of Tokyo, Bunkyo-ku, Tokyo 113-0033, Japan}


\begin{abstract}
Weakly interacting massive particles provide a well-motivated framework for dark matter, naturally reproducing the observed relic abundance through thermal freeze-out. A recent claim of an indirect-detection signal from the Galactic halo, consistent with dark matter annihilation in the mass range $400$--$800~\mathrm{GeV}$, motivates a reexamination of minimal models that can account for such a signal while remaining consistent with existing constraints. In this paper, we analyze the simplest extensions of the Standard Model capable of explaining the signal. We show that electroweak doublet dark matter with Higgs-portal interactions provides a natural and economical explanation. The model predicts annihilation predominantly into gauge bosons with characteristic branching fractions and allows for inelastic dark matter with a mass splitting of order $100~\mathrm{keV}$, intriguingly consistent with a recent direct-detection anomaly. Possible enhancements of the present-day annihilation rate relative to the thermal value are also discussed, including a simple extension with a light scalar field that induces Sommerfeld enhancement.
\end{abstract}

\maketitle

\section{Introduction}
\label{sec:intro}

Identifying the particle nature of dark matter remains one of the central problems in particle physics and cosmology. Among the many possibilities, weakly interacting massive particles (WIMPs) stand out as a particularly well-motivated class of candidates. A WIMP with electroweak-scale mass and interactions naturally reproduces the observed dark matter relic abundance through thermal freeze-out~\cite{Lee:1977ua,Steigman:2012nb}
\begin{equation}
  \langle \sigma v \rangle_{\rm th} \sim (2\text{--}4)\times 10^{-26}~\mathrm{cm^3/s},
\label{eq:sigma-th}
\end{equation}
a coincidence often referred to as the ``WIMP miracle.''

Indirect-detection experiments provide a complementary probe of WIMPs by searching for Standard Model particles produced by dark matter annihilation in astrophysical environments. The Galactic Center is a particularly promising target due to its large dark matter density, albeit with significant astrophysical uncertainties.

Recently, Ref.~\cite{Totani:2025fxx} reported a possible excess in indirect-detection data from the Galactic halo, spatially distinct from the previously reported Galactic Center excess~\cite{Daylan:2014rsa,Fermi-LAT:2017opo}, and characterized by a significantly higher photon-energy peak (at $\sim 20~\mathrm{GeV}$, compared to $\sim 2~\mathrm{GeV}$ for the Galactic Center excess). If interpreted as dark matter annihilation, the signal favors a mass range
\begin{equation}
  m_{\rm DM} \sim 400\text{--}800~\mathrm{GeV},
\end{equation}
with an annihilation cross section close to the canonical thermal value, possibly larger by one or two orders of magnitude depending on assumptions about astrophysical boost factors.

In this paper, we address the following question: \emph{what is the simplest particle-physics model that can account for this signal while remaining consistent with existing experimental constraints?} Our guiding principles are:
\begin{itemize}
\item minimal field content with renormalizable interactions,
\item a thermal freeze-out origin of the relic abundance,
\item consistency with relic abundance, direct detection, collider bounds, and indirect detection.
\end{itemize}
While the dominant annihilation modes considered in Ref.~\cite{Totani:2025fxx} were $b\bar b$ and $W^+W^-$, we find that other heavy Standard Model final states provide equally good descriptions of the data. This observation motivates us to consider scenarios in which dark matter couples to the Higgs sector, with the simplest realization given by a singlet scalar Higgs-portal model. However, this possibility is ruled out by direct-detection experiments, which leads us to consider electroweakly charged dark matter, in particular a second Higgs doublet whose neutral component constitutes the dark matter~\cite{Deshpande:1977rw,Barbieri:2006dq}. We study both the minimal realization of this scenario and a simple extension involving an additional scalar degree of freedom.

We find that dark matter in these models is naturally inelastic~\cite{Tucker-Smith:2001myb}, with a mass splitting consistent with recently suggested direct-detection anomalies~\cite{An:2025bby}. The scenario places dark matter in a mass range accessible to indirect detection, direct detection, and future collider experiments, offering a rare opportunity for a coherent experimental test.

\section{Fits to Heavy Final States}
\label{sec:fits}

\begin{figure}[t]
\centering
\includegraphics[width=\linewidth]{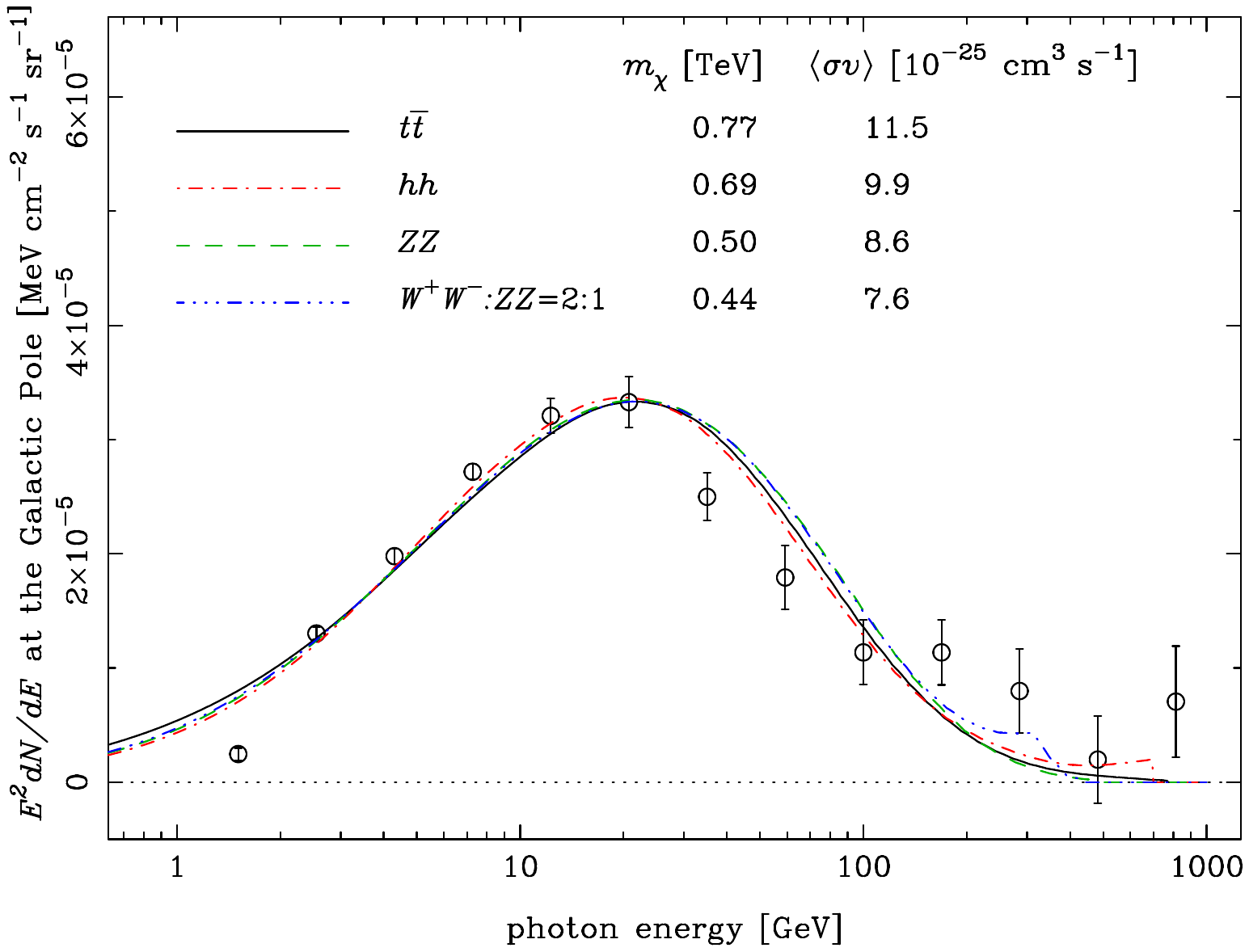}
\caption{
 Representative fits to the Galactic halo signal for dark matter annihilation into $t\bar t$, $hh$, and $ZZ$ final states, as well as electroweak gauge-boson final states with $W^+ W^- : Z Z = 2 : 1$ as in Eq.~(\ref{eq:gauge-final}). The best-fit dark matter mass and annihilation cross section for each channel are indicated in the legend. Masses in the range $m_{\rm DM} \sim 400$--$800~\mathrm{GeV}$ provide comparably good fits. The data points are the fit using the supernova remnant (SNR) distribution for cosmic-ray sources, shown in Fig.~16 of Ref.~\cite{Totani:2025fxx}.
}
\label{fig:fits}
\end{figure}

We first note that the Galactic halo signal can be fit not only by $b\bar b$ and $W^+W^-$ final states, but also by
\begin{equation}
  t\bar t,\qquad hh,\qquad ZZ,
\end{equation}
with comparable quality in the mass range $m_{\rm DM}\sim 400$--$800~\mathrm{GeV}$. Figure~\ref{fig:fits} illustrates representative fits for these channels. The best-fit annihilation cross section and mass vary mildly across channels, but the former remains close to the canonical thermal value, modulo astrophysical uncertainties, while the latter lies in a common range of a few hundred GeV.

The fact that these additional heavy final states also provide good fits is important for model building. In simple renormalizable completions, an annihilation channel such as $b\bar b$ or $W^+W^-$ rarely appears in isolation. For example, if dark matter annihilates through couplings to the Higgs sector, then in the mass range of interest the leading final states are generically $W^+W^-$, $ZZ$, and $hh$, with $t\bar t$ and $b\bar b$ subdominant. Likewise, if the annihilation is governed by electroweak gauge interactions, a $W^+W^-$ channel is typically accompanied by a significant $ZZ$ contribution, while fermionic final states are suppressed. The fact that $t\bar t$, $hh$, and $ZZ$ also fit the data therefore implies that these unavoidable accompanying channels do not spoil the phenomenological interpretation, significantly broadening the class of simple viable models.

A natural realization of gauge-dominated annihilation is provided by electroweakly charged fermions, such as a Majorana $SU(2)_L$ triplet (wino-like) or a Higgsino-like fermion. However, for these candidates the relic abundance is fixed by gauge interactions, leading to a mass of order a few TeV for a triplet and $\sim 1~\mathrm{TeV}$ for a doublet. These values lie well above the mass range suggested by the signal. Therefore, such scenarios do not provide a simple explanation of the observed excess within the framework of thermal relic dark matter with renormalizable interactions.

\section{Singlet Higgs-Portal Dark Matter}
\label{sec:singlet}

We now turn to the possibility that dark matter annihilates through couplings to the Higgs sector. The simplest realization is a real scalar $\phi$, stabilized by a $\mathbb{Z}_2$ symmetry $\phi\to-\phi$, with interaction
\begin{equation}
  \mathcal{L} \supset -\frac{\lambda}{2}\,H^\dagger H\,\phi^2.
\end{equation}

In the heavy-mass limit $m_\phi \gg m_W,m_Z,m_h$, annihilation proceeds dominantly into gauge and Higgs bosons, with the longitudinal modes of the gauge bosons dominating in this limit,
\begin{equation}
  \phi\phi \to W_L^+ W_L^-,\ Z_L Z_L,\ hh,
\end{equation}
with approximate branching ratios
\begin{equation}
  W_L^+ W_L^- : Z_L Z_L : hh \simeq 2:1:1.
\end{equation}
The contribution to $t\bar t$ is suppressed relative to the bosonic channels by $m_t^2/m_\phi^2$ and is typically at the $10\%$ level or below.

Fixing $\lambda$ by the thermal relic abundance leads to
\begin{equation}
  \lambda \simeq 0.15 \left( \frac{m_\phi}{500~\mathrm{GeV}} \right),
\end{equation}
which is capable of reproducing the indirect-detection signal. However, the same coupling induces elastic scattering on nuclei via Higgs exchange, leading to spin-independent cross sections well above current experimental bounds~\cite{LZ:2022lsv,XENON:2023cxc}. The conclusion is unchanged if $\phi$ is taken to be complex. This model is therefore excluded over the parameter space of interest.

\section{Electroweakly Charged Higgs-Portal Dark Matter}
\label{sec:EW-doublet}

We now turn to the possibility that dark matter is charged under the electroweak gauge group. Motivated by the dominance of $W^+W^-$ and $ZZ$ final states, we consider a scalar field $\Phi$ transforming under the Standard Model gauge group as
\begin{equation}
  \Phi \sim (\mathbf{1},\mathbf{2})_{1/2},
\end{equation}
i.e.\ with the same quantum numbers as the Standard Model Higgs field.%
\footnote{
 A real scalar electroweak triplet $(\mathbf{1},\mathbf{3})_0$ provides another example of gauge-dominated annihilation without tree-level $Z$ exchange. However, its thermal relic mass is typically of order a few TeV, well above the mass range of interest here, making it less suitable for explaining the observed signal.
}

In this case, annihilation is dominated by electroweak interactions, with contributions from both gauge interactions and scalar quartic couplings. In the heavy-mass limit, it proceeds as
\begin{equation}
  \Phi\Phi \to W^+ W^-,\ Z Z,
\end{equation}
with approximate branching ratios
\begin{equation}
  W^+ W^- : Z Z \simeq 2:1.
\label{eq:gauge-final}
\end{equation}
In the regime where annihilation is dominated by electroweak gauge interactions, the produced gauge bosons are predominantly transversely polarized. The annihilation cross section scales approximately as
\begin{equation}
  \langle \sigma v \rangle_{\rm th} \sim \frac{g^4}{\mathcal{O}(10)\,\pi m_\Phi^2}.
\label{eq:sigma-gauge}
\end{equation}
Assuming that the lightest component of $\Phi$ is a neutral state and constitutes all dark matter, this leads to~\cite{Banerjee:2019luv,Justino:2024etz}
\begin{equation}
  m_{\rm DM} \sim 500\text{--}600~\mathrm{GeV}.
\end{equation}
This relatively narrow mass range follows from the fact that the annihilation cross section in Eq.~(\ref{eq:sigma-gauge}) is determined predominantly by electroweak gauge interactions, with only mild dependence on additional parameters. As a result, requiring the correct relic abundance fixes $m_{\rm DM}$ to be in the few-hundred-GeV range, consistent with the indirect-detection signal.%
\footnote{
Early analyses of the inert doublet model found that for $m_{\rm DM} \gtrsim m_W$ the relic abundance is typically underproduced~\cite{Barbieri:2006dq}. This corresponds to a different parameter regime, motivated by a heavy Standard Model Higgs, in which scalar quartic couplings give sizable contributions to annihilation into longitudinal gauge bosons. In the regime considered here, these contributions are not comparably enhanced, so that annihilation is controlled primarily by electroweak interactions as we will see below.
}

The regime described above is realized when the contribution of scalar quartic interactions to the annihilation amplitude is not dominant. To make this discussion precise, we write the most general renormalizable scalar interactions consistent with the $\Phi \to -\Phi$ symmetry:
\begin{align}
  V_{\rm int}(H,\Phi) ={}& \lambda_1 (H^\dagger H)^2 + \lambda_2 (\Phi^\dagger \Phi)^2
\nonumber\\
  &+ \lambda_3 (H^\dagger H)(\Phi^\dagger \Phi)
  + \lambda_4 (H^\dagger \Phi)(\Phi^\dagger H)
\nonumber\\
  &+ \frac{\lambda_5}{2}\Bigl[(H^\dagger \Phi)^2 + {\rm h.c.}\Bigr].
\end{align}
After electroweak symmetry breaking, the neutral components of $\Phi$ split into two real scalar states with a mass difference controlled by $|\lambda_5|$. The coupling of the dark matter state to the Higgs boson is governed by the combination
\begin{equation}
  \lambda_L \equiv \lambda_3 + \lambda_4 + {\rm Re}\,\lambda_5 .
\label{eq:lambda_L}
\end{equation}

Direct detection through Higgs exchange is governed by $\lambda_L$, and current bounds require this combination to be small, typically $|\lambda_L| \lesssim 10^{-3}\text{--}10^{-2}$. However, annihilation into $W_L^+ W_L^-$, $Z_L Z_L$, and $hh$ also receives contributions from the scalar quartic couplings separately. In principle, sizable scalar-potential contributions to annihilation may coexist with a small $\lambda_L$ through cancellations among different couplings. In this work, we focus on the simpler regime in which such cancellations are not important, so that the relic abundance is controlled primarily by electroweak interactions.

A neutral electroweak doublet is naively excluded by direct-detection constraints due to tree-level $Z$ exchange. This leads to large elastic scattering cross sections with nuclei, far above experimental bounds. This problem can be avoided if the dark matter is inelastic~\cite{Tucker-Smith:2001myb}. With general renormalizable couplings between $H$ and $\Phi$, the two neutral components split with a mass difference
\begin{equation}
  \Delta m \sim |\lambda_5|\,\frac{\langle H \rangle^2}{m_{\rm DM}}.
\label{eq:delta-m}
\end{equation}
If $\Delta m \gtrsim m_{\rm DM} v^2$ with $v\sim 10^{-3}$, elastic scattering is kinematically forbidden, and the model is consistent with direct-detection constraints. This construction is known as inert doublet dark matter~\cite{Barbieri:2006dq}.

In Fig.~\ref{fig:fits}, we show representative fits to the Galactic halo signal for the electroweak gauge-boson final states in Eq.~(\ref{eq:gauge-final}), together with other representative channels. The best-fit dark matter mass is
\begin{equation}
  m_{\rm DM} \simeq 440~{\rm GeV},
\end{equation}
which lies within the mass range suggested by thermal freeze-out in this model. We note that the inferred value of $m_{\rm DM}$ is subject to systematic uncertainties associated with the data analysis. In Ref.~\cite{Totani:2025fxx}, different fitting procedures applied to the same NFW-$\rho^2$ model lead to variations in the best-fit mass by a factor of $\sim 1.7$. We also note that the fit is not sharply peaked, and masses in the range $m_{\rm DM} \sim 400$--$800~\mathrm{GeV}$ provide comparably good descriptions of the data.

The corresponding annihilation cross section is
\begin{equation}
  \langle \sigma v \rangle \simeq 8 \times 10^{-25}~\mathrm{cm^3/s}.
\label{eq:ann-ID}
\end{equation}
This value exceeds the canonical thermal cross section in Eq.~(\ref{eq:sigma-th}) by a factor of $\mathcal{O}(10)$. This discrepancy may be accounted for by astrophysical boost factors and/or by additional particle-physics effects that enhance the present-day annihilation rate relative to that at freeze-out.

\section{Connection to a Direct-Detection Anomaly}
\label{sec:anomaly}

The inelastic splitting between the two neutral components is given parametrically by Eq.~(\ref{eq:delta-m}). In order to evade direct-detection constraints from inelastic upscattering, the splitting must exceed the typical kinetic energy available in halo dark matter scattering,
\begin{equation}
  \Delta m \gtrsim m_{\rm DM} v^2 \sim \mathcal{O}(100~{\rm keV}),
\end{equation}
for $m_{\rm DM} \sim 500~{\rm GeV}$ and $v \sim 10^{-3}$. This corresponds to a lower bound on the coupling
\begin{equation}
  |\lambda_5| \gtrsim 10^{-6}.
\end{equation}

A small value of $|\lambda_5|$ is technically natural. In the limit $\lambda_5 \to 0$, the scalar potential acquires an enhanced global $U(1)$ symmetry acting on the inert doublet $\Phi$, under which $\Phi \to e^{i\alpha}\Phi$. The term proportional to $\lambda_5$ is the only renormalizable interaction in the scalar potential that breaks this symmetry. Therefore, taking $|\lambda_5| \ll 1$ is stable against radiative corrections.

Remarkably, Ref.~\cite{An:2025bby} reports an excess consistent with precisely this range of masses ($\sim 500~{\rm GeV}$) and splittings ($\sim 150~{\rm keV}$), which, using Eq.~(\ref{eq:delta-m}), corresponds to $|\lambda_5| \sim 10^{-6}$, providing an intriguing possible connection.

\section{Enhancing Indirect Detection}
\label{sec:enhancing}

So far we have assumed that the indirect-detection signal is compatible with the thermal annihilation rate due to astrophysical effects such as boost factors. If this assumption fails, a dynamical enhancement of the present-day annihilation rate may be required.

To enhance the present-day annihilation rate relative to that at freeze-out, one may introduce a light scalar field $\Sigma$ coupled to the dark matter field $\Phi$ through
\begin{equation}
  \mathcal{L} \supset \mu\,\Sigma\,\Phi^\dagger\Phi,
\end{equation}
where $\mu$ is a dimensionful coupling. Exchange of $\Sigma$ between two nonrelativistic dark matter particles generates an attractive Yukawa potential
\begin{equation}
  V(r) \simeq -\,\frac{\alpha_\Sigma}{r}\,e^{-m_\Sigma r},
\end{equation}
with an effective coupling
\begin{equation}
  \alpha_\Sigma \sim \frac{\mu^2}{16\pi m_{\rm DM}^2},
\end{equation}
up to factors of order unity.

Sommerfeld enhancement~\cite{Hisano:2004ds} becomes significant when the interaction range is comparable to or larger than the de Broglie wavelength of halo dark matter, i.e.
\begin{equation}
  m_\Sigma \lesssim m_{\rm DM} v.
\end{equation}
For the mass range of interest in this work, $m_{\rm DM}\sim 500~{\rm GeV}$, this condition becomes
\begin{equation}
  m_\Sigma \lesssim 500~{\rm MeV} \left(\frac{m_{\rm DM}}{500~{\rm GeV}}\right) \left(\frac{v}{10^{-3}}\right),
\label{eq:upper}
\end{equation}
for Galactic halo annihilation. In the Coulomb limit, $m_\Sigma \ll m_{\rm DM} v$, the enhancement factor is approximately
\begin{equation}
  S \simeq \frac{\pi \alpha_\Sigma/v}{1-e^{-\pi \alpha_\Sigma/v}},
\end{equation}
which reduces to $S \simeq \pi\alpha_\Sigma/v$ for $\alpha_\Sigma \gg v$.

To account for the discrepancy between Eq.~(\ref{eq:sigma-th}) and Eq.~(\ref{eq:ann-ID}), one may take, for example,
\begin{equation}
  \frac{\alpha_\Sigma}{v} \sim 10,
\end{equation}
which implies
\begin{equation}
  \mu \sim 0.7\, m_{\rm DM}
  \sim 350~{\rm GeV}
  \left(\frac{m_{\rm DM}}{500~{\rm GeV}}\right).
\end{equation}
Larger values of $\mu$ and smaller values of $m_\Sigma$ can lead to substantially stronger enhancement, and resonant enhancement is also possible for special values of $\alpha_\Sigma m_{\rm DM}/m_\Sigma$.

The annihilation strength in Eq.~(\ref{eq:ann-ID}) has a mild tension with the upper bound from dwarf spheroidal galaxies~\cite{McDaniel:2023bju}. While this level of tension may well be resolved by astrophysical uncertainties, this would require that the Sommerfeld enhancement has already saturated at velocities well above those characteristic of dwarf spheroidal galaxies. In particular, since dwarf spheroidal galaxies have a typical dark matter velocity $v \simeq 5 \times 10^{-5}$, the condition
\begin{equation}
  m_\Sigma \gg m_{\rm DM} v \sim 25~{\rm MeV}
  \left(\frac{m_{\rm DM}}{500~{\rm GeV}}\right)
  \left(\frac{v}{5 \times 10^{-5}}\right)
\end{equation}
ensures that the Sommerfeld enhancement has already saturated in these systems, so that the annihilation cross section is not further enhanced relative to that in the Galactic halo. Together with the upper bound in Eq.~(\ref{eq:upper}), this requirement delineates a parametric window for $m_\Sigma$.

The introduction of the scalar field $\Sigma$ does not qualitatively affect the discussion above, provided that its couplings to the Standard Model Higgs field are sufficiently small. In this case, mixing with the Higgs boson is negligible, and the relic abundance and direct-detection constraints remain essentially unchanged. Collider phenomenology is also largely unaffected, although additional rare processes may arise. For example, if $\Sigma$ couples to the Higgs field via a term $\mu' \Sigma H^\dagger H$, together with self-interactions of $\Sigma$, invisible decays of the Higgs boson can be induced, with branching ratios parametrically of order $(\mu'/m_h)^2$.

Recent works have explored alternative mechanisms to enhance the present-day annihilation rate. Ref.~\cite{Murayama:2025ihg} considers resonant annihilation with $m_\Sigma \simeq 2 m_{\rm DM}$, while Ref.~\cite{Jho:2025iah} studies Sommerfeld enhancement in the presence of $p$-wave annihilation. The model presented in Ref.~\cite{Murayama:2025ihg} relies on Higgs-portal interactions to set the relic abundance, and is therefore subject to the direct-detection constraints discussed in Section~\ref{sec:singlet}.

\section{Summary and Outlook}
\label{sec:summary}

We have shown that a minimal inert-doublet extension of the Standard Model provides a simple and compelling explanation of the Galactic halo gamma-ray excess reported in Ref.~\cite{Totani:2025fxx}. In this framework, the relic abundance is determined primarily by electroweak gauge interactions, while Higgs-portal couplings are sufficiently suppressed to evade direct-detection constraints. The model naturally predicts dominant annihilation into $W$ and $Z$ bosons in the mass range $400$--$800~\mathrm{GeV}$, in agreement with the observed signal.

We have also considered a simple extension of this framework involving a light scalar field that induces Sommerfeld enhancement of the present-day annihilation rate. The scalar mass is constrained to lie in a parametric window in which the enhancement is effective in the Galactic halo while already saturated at the characteristic velocities of dwarf spheroidal galaxies.

The inelastic structure of the neutral states suppresses elastic scattering through $Z$ exchange, allowing the model to satisfy stringent direct-detection bounds. Remarkably, the mass scale and splitting suggested by a recent direct-detection anomaly are consistent with the parametric expectations of this scenario.

Collider constraints on the electroweak doublet sector are relatively weak in the parameter region of interest. LEP bounds on charged scalars are easily satisfied, while LHC searches for electroweak states have reduced sensitivity for masses of order several hundred GeV, especially in the presence of small mass splittings. As a result, the parameter space considered here remains consistent with current collider limits. A future lepton collider with $\sqrt{s}\gtrsim 2m_{\rm DM}$, corresponding to center-of-mass energies below $\sim 2~\mathrm{TeV}$ for the masses of interest, would provide a decisive test of this scenario.

Dark matter in this framework lies at the intersection of indirect detection, direct detection, and collider searches, offering a rare opportunity for a coherent experimental probe across multiple frontiers.

\acknowledgments

We thank Dan Kondo and Satoshi Shirai for useful discussions. This work was initiated through conversations between Y.N.\ and T.T.\ during their joint appearance on the YouTube channel ``ReHacQ'' (https://youtu.be/6ZBE5fQAZXM and https://youtu.be/p30fgjFrAtY), and we gratefully acknowledge this opportunity.

The work of Y.N.\ was supported by the U.S. Department of Energy, Office of Science, Office of High Energy Physics under QuantISED Award DE-SC0019380 and Contract No.\ DE-AC02-05CH11231, by MEXT KAKENHI Grant No.\ JP25K00997, and by the Japan Science and Technology Agency (JST) as part of Adopting Sustainable Partnerships for Innovative Research Ecosystem (ASPIRE), Grant No.\ JPMJAP2318. The work of T.T.\ was supported by JSPS/MEXT KAKENHI Grant No.\ 18K03692, 26H00822, and 26H02063.

\end{document}